\documentclass[conference]{IEEEtran}
\IEEEoverridecommandlockouts
\usepackage{cite}
\usepackage{amsmath,amssymb,amsfonts}
\usepackage{algorithmic}
\usepackage{graphicx}
\usepackage{multirow}
\usepackage{textcomp}
\usepackage{xcolor}
\usepackage{svg} 

\graphicspath{{./pic/}}

\def\BibTeX{{\rm B\kern-.05em{\sc i\kern-.025em b}\kern-.08em
    T\kern-.1667em\lower.7ex\hbox{E}\kern-.125emX}}
\begin{document}

\title{Review of Ansatz Designing Techniques for Variational Quantum Algorithms
}

\author{\IEEEauthorblockN{Junhan Qin}
\IEEEauthorblockA{\textit{School of Information Engineering} \\
\textit{Minzu University of China}\\
Beijing, China \\
19011411@muc.edu.cn}
}

\maketitle

\begin{abstract}
For a large number of tasks, quantum computing demonstrates the potential for exponential acceleration over classical computing. In the NISQ era, variable-component subcircuits enable applications of quantum computing. To reduce the inherent noise and qubit size limitations of quantum computers, existing research has improved the accuracy and efficiency of Variational Quantum Algorithm (VQA). In this paper, we explore the various ansatz improvement methods for VQAs at the gate level and pulse level, and classify, evaluate and summarize them.
\end{abstract}


\section{Introduction}
Quantum computing (QC) is a new computing paradigm with great potential for exponential acceleration beyond classical computers to solve classically intractable problems with greater efficiency and speed. In many fields, such as pattern recognition~\cite{RuanYue2014Quantum} and data mining~\cite{Wittek2014Quantum}, it has advantages over traditional computers. In the past two decades, quantum hardware has also made breakthroughs in physical realization technology and made great progress, which provides the possibility for the application of quantum computing.

Although Quantum computing has made considerable progress, we 
should be aware that we are still in the Noisy Intermediate Scale
Quantum (NISQ)~\cite{JohnPreskill2018QuantumCI} stage. Currently,
quantum computers usually only have tens to hundreds of qubits. 
Limited by the hardware equipment, it is impossible to build a 
universal fault-tolerant quantum computer. However, the emergence 
of variational quantum algorithms 
(VQA)~\cite{MarcoCerezo2021VariationalQA} makes the current 
application of quantum computing possible. We can construct 
different quantum circuits and loss functions for different 
problems. Common VQAs are variational quantum eigensolver 
(VQE)~\cite{DmitryAFedorov2021VQEMA, HanruiWang2021QuantumNASNS, meitei2021gate}, dynamical quantum 
simulation~\cite{XueyiGuo2018ObservationOD}, and quantum 
approximation optimization algorithm 
(QAOA)~\cite{EdwardFarhi2016QuantumST,liang2022hybrid, guerreschi2019qaoa}, quantum machine learning (QML)~\cite{JacobBiamonte2017QuantumML,jiang2021machine,wang2021exploration,liang2022variational,qi2022classical,qin2022improving} and so on.

However, VQA still faces many difficulties. Since the current VQA is mostly based on the quantum environment simulated by the classical simulators, once applied to the real machine, The huge noise of NISQ phase quantum devices (The qubits and quantum gates suffer from high error rates of $10^{-3}$ to $10^{-2}$) will significantly affect the accuracy of the algorithm~\cite{jiang2021machine, wang2021exploration}. Similarly, limited qubits also limit the depth of the algorithm, affecting the efficiency and accuracy of the algorithm.

This article summarizes existing VQA improvement approaches and outlines the different directions for improvement at the gate level and pulse level respectively. We will first look at improved algorithms at the gate level, As shown in the following table, VQA methods can be divided into two categories: improving accuracy and improving efficiency. Furthermore, on this basis,  the two directions will adopt different methods to achieve the goal. After a brief introduction, this paper will discuss whether it has been successfully applied to real machines, whether it is device dependent, and make a summary of different methods. For the pulse level, we found that the current review is not covered much, but it is a promising direction, and we will select representative methods to brief and evaluate.

The rest of this article is organized as follows. Section II Outlines the basics of quantum computing. In the same section, the motivation of this paper will be briefly discussed. Section III will introduce different methods implemented to improve VQAs. Finally, Section IV will conclude and discuss some perspective open questions.

\section{BACKGROUND AND MOTIVATION}

\subsection{Quantum Basics}

\textbf{Qubit.}
Unlike conventional bits, which can represent only 0 and 1 states, 
a quantum bit (qubit) can represent various linear combinations of the two basic states, 0 and 1:
$|\phi \rangle$ = $\alpha\lvert0\rangle + \beta\lvert1\rangle$, 
for $\alpha,\beta \in \mathbb{C}$, 
$ \lvert a \rvert ^2 +  \lvert b \rvert ^2$ = 1, this is also been called superposition. Because of this unique restriction, an N-qubit system can represent $2^n$ states, 
whereas conventional bits can represent only one state at a time. 
The qubit of a quantum computer can generally be simulated in the following ways:  
1. An electron orbiting a hydrogen atom may be in the ground state($\lvert0\rangle$) or activated state($\lvert1\rangle$);
2. An electron's spin may be up ($\lvert0\rangle$) or down ($\lvert1\rangle$); 
3. Photon Polarization direction, may be horizontal ($\lvert0\rangle$), or along the vertical ($\lvert1\rangle$);

\textbf{Quantum gate and Quantum circuit..}
In classical computation, all of the logical operations can be represented with "and", "or", and "not" three operations. In quantum computing, we also need a similar scheme to complete the corresponding operation,
which is the quantum gate. In the computational model of quantum computing, especially quantum circuits, 
a quantum gate is the basic component of a quantum circuit that operates on the Pauli operations. The relationship between a quantum gate and a quantum circuit is similar to the relationship between conventional logic gates and conventional digital
circuits. Quantum gates are often represented by unitary matrices, and a gate that operates on K qubits can be
represented by a 2K x 2k unitary matrix.
The operation of a quantum gate can be represented by multiplying a matrix representing the gate with a
vector representing the state of the qubit. 
Common quantum gates are divided into single-qubit gates and multi-qubit gates. 
Single qubit gates include Hadamard Gate, Pauli Gate, and so on. 
Multi-qubit gates have Controlled - Not gate (CX gate)+, Controlled Phase gate (CZ gate), and SWAP gate.

To use a quantum computer to perform calculations, we need to build a quantum
circuit to manipulate the state of the qubit. A quantum circuit consists of multiple quantum
gates. 

\textbf{Variational Quantum Algorithm (VQA).}
The variational quantum algorithms are to train a quantum circuit with a classical optimizer. The goal of a VQA is to find a model that can fit the data, that is, to determine an optimal set of parameters to make the model as close as possible to the given data. The algorithm of VQA includes three parts: constructing a cost function, building an ansatz, and adjusting parameters. The common VQA includes variational quantum eigensolver (VQE), quantum approximation optimization algorithm (QAOA), quantum machine learning (QML), and so on. VQA is a promising direction, but it still faces some difficulties. The first is the trainability of quantum circuits. In some cases, such as barren plateau, the optimizer cannot effectively train the large circuit due to the limited decoherence time. The noise of quantum devices can affect the accuracy of VQA in many ways. Physical noise may slow down the training process, making the noisy global optimum no longer correspond to the noiseless global optimum, and thus affecting the final optimal loss value.

\begin{figure}[htbp]
	\centering
	\includegraphics[width=\linewidth]{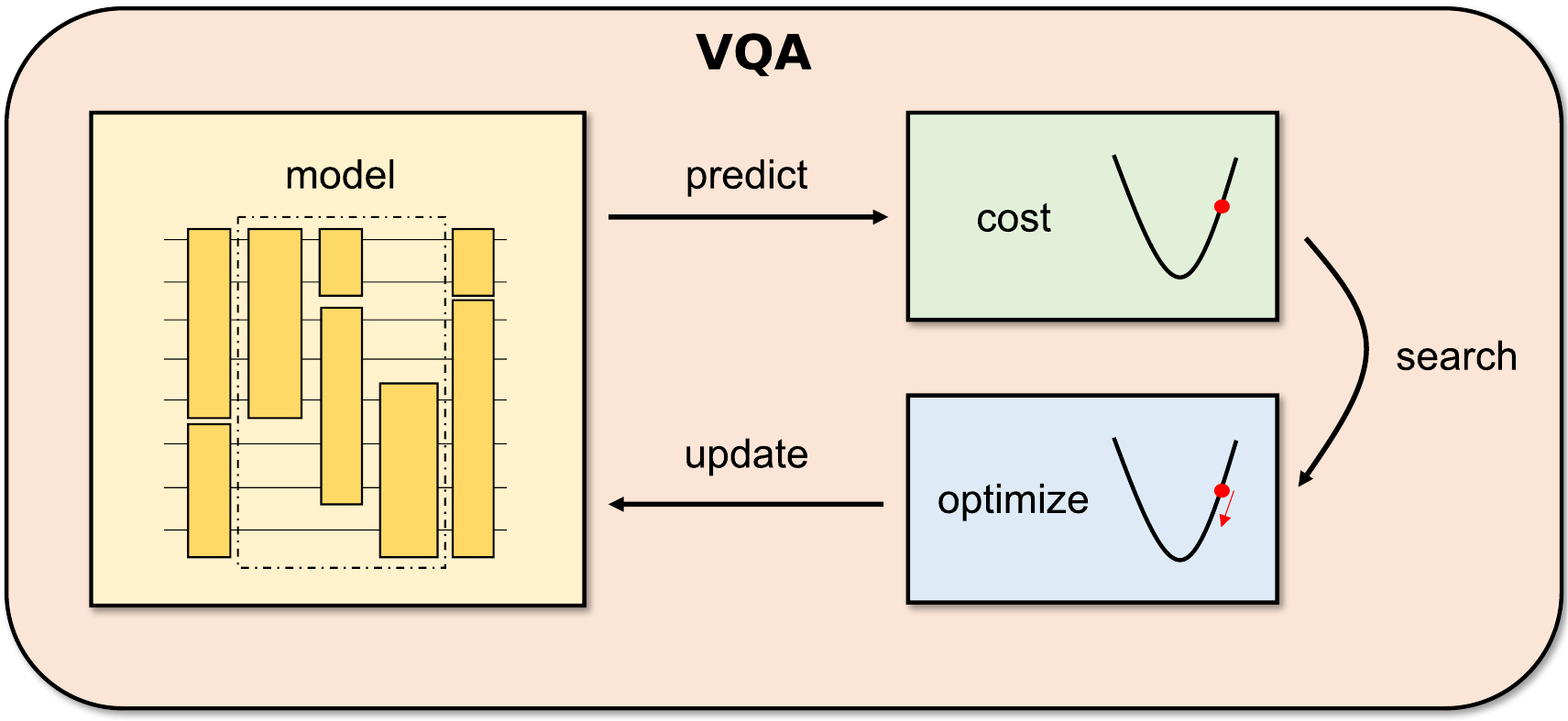}
	\caption{Schematic of variational quantum algorithm, the model is designed based on quantum gates on quantum computer, and the optimization progress is on classical computer, the classical computer optimizes and updates the parameters in the trainable layer of the model.}
	\label{fig:pic1}
\end{figure}

\textbf{Variational Quantum Eigensolver (VQE).}
VQE is an algorithm that uses classical optimizers to train a parametric quantum circuit for solving matrix eigenvalues and eigenvectors, it is the earliest variational quantum algorithm and is mainly used to solve the ground state and low excited state of the quantum system. Due to this property, VQE has been widely used to solve problems in quantum chemistry. For a quantum system, the dimension of the Hamiltonian $\mathbb{H}$ grows exponentially with the system size. The other algorithms often require millions of qubits, far beyond the reach of current quantum computers. Thus, VQE was proposed to work with NISQ machines.

\textbf{Quantum Machine Learning.}
A quantum algorithm can perform quantum supremacy in solving specific tasks than the algorithms of the fastest known classical computers.
N-dimensional data can theoretically be represented by only log(N) bits in qubits, 
unlike classical computer algorithms which require N bits. Therefore, people can design various algorithms
to achieve acceleration by quantum computers, which is quantum machine learning. Some of the quantum machine learning inspried by classical machine learning and produced some useful quantum algorithms, such as qPCA~\cite{huang2022quantum,lloyd2014quantum}, qSVM~\cite{rebentrost2014quantum, li2015experimental, ding2021quantum}, quantum deep learning~\cite{wiebe2014quantum,levine2019quantum} and so on.
However, due to the relatively small scale of the existing quantum computers, and the relatively deep size
and depth required by the existing algorithms, there is still a great gap to truly surpass classical machine learning. In addition, the noise problem of the quantum computer, barren plateau phenomenon, etc. in the noise-intermediate-scale quantum (NISQ) era machines, also restrict the performance of quantum machine learning.

\textbf{Noise on the NISQ machine.}
First, the noise in quantum computing, due to the imperfect control signal crosstalk between qubits, real quantum computer interaction, and interference from the external environment, inevitably there will be noise, this greatly affected the accuracy of quantum machine learning, so you need to frequent the quantum machine were characterized and calibration, to reduce the influence of noise. 
At the same time, NISQ devices incur decoherence errors over time, which limited the depth and width of the circuit, making it impractical to implement complex quantum circuits.

\textbf{Ansatz.}
ansatz is the basic architecture of a circuit, a set of gates that act on a specific subsystem \cite{cheng2022topgen}.
ansatz is essentially a few prior assumptions, which are guesses about the appropriate training circuit. On the one hand, the problem ansatz usually depends on the specific problem, it can combine the property of the problem and then solve the problem. On the other hand, some ansatz architectures are generic and problem-neutral, meaning that they can be used generally for kinds of problems on quantum hardware, we call this type of ansatz as hardware ansatz.

\subsection{Motivation}
In recent years, the research community on quantum computing has gradually grown. More and more studies show that compared with classical computers, VQAs have excellent acceleration potential in a large number of tasks. Studies have shown that a two-dimensional programmable superconducting quantum processor, Zuchongzhi, which is composed of 66 functional qubits, can do a random quantum circuit sampling task in 1.2 hours,  which would take eight years for today's most powerful supercomputers~\cite{YulinWu2021StrongQC}. The advantage of quantum computing is that an n-dimensional piece of data (a vector) can theoretically be represented by only log(N) bits, rather than the classical N bits. One qubit can be in a superposition of 0 and 1 states and be coherent with each other, and multiple qubits can be entangled to represent more complex states or perform more complex operations. 

However, the current VQAs still face great challenges. Most of the advantages of existing VQAs are demonstrated in the quantum environment simulated by the classical simulators, and their efficiency and accuracy will decrease significantly in the real quantum environment in NISQ machines. This is due to the high noise characteristics of NISQ machines and the small scale of qubits of existing quantum computers.  In the case that the scale and technology of quantum computers still have a gap with quantum supremacy, what we can do in the gate level optimization is to further optimize the quantum circuit to improve the efficiency and accuracy to reduce the gap. Furthermore, some researchers state the gate level is not a good abstraction layer, and several works have been proposed to gain the benefits from pulse level control.

To better understand the function of state of art circuit optimization methods as well as make it easier for choosing the matched framework for users' quantum program. We proposed a review of existing circuit optimization methods for VQAs in Section \ref{sec3}. We can find a variety of optimization methods and whether it is useful. In terms of methods, some studies focus on reducing the impact of noise, some use pruning to reduce overhead, and some take a holistic view and use some algorithms of traditional machine learning for circuit optimization. At the same time, there will be a discussion on whether it has been successfully applied to real machines and whether it is device dependent.

\begin{table*}[t]
\caption{Proposed category of the summarized existing techniques. Three major categories and five sub-categories were defined and the models involved in this review stated, besides the application-specific or not for each model discussed.}
\centering
\renewcommand*{\arraystretch}{1}
\setlength{\tabcolsep}{10pt}
\footnotesize
\begin{tabular}{|cc|c|c|}
\hline
\multicolumn{1}{|c|}{Method} & { Specific Method } & {Model} & { Application Specific or general } \\ \hline
\multicolumn{1}{|c|}{\multirow{7}{*}{ Improving Accuracy }} & \multirow{2}{*}{Noise Suppression} & { QuantumNAT~\cite{HanruiWang2021QuantumNATQN} } & General   \\ \cline{3-4} 
\multicolumn{1}{|c|}{} &  & { QF-RobustNN~\cite{ZhidingLiang2021CanNO} } & Specific  \\ \cline{2-4} 
\multicolumn{1}{|c|}{} & \multirow{4}{*}{Better Circuit} & QAS~\cite{YuxuanDu2020QuantumCA} & General  \\ \cline{3-4} 
\multicolumn{1}{|c|}{} &  & { QuantumNAS~\cite{HanruiWang2021QuantumNASNS} } & General  \\ \cline{3-4} 
\multicolumn{1}{|c|}{} &  &  { QMLP~\cite{ChengChu2022QMLPAE} } & Specific  \\ \cline{3-4} 
\multicolumn{1}{|c|}{} &  & { DRL-QAS~\cite{EnJuiKuo2021QuantumAS} } & General  \\ \cline{2-4} 
\multicolumn{1}{|c|}{} &{ Approaches for particular problems } & QUILT~\cite{DanielSilver2022QUILTEM} & Specific  \\ \hline
\multicolumn{1}{|c|}{\multirow{4}{*}{Improving Efficiency}} & \multirow{2}{*}{Pruning} & On-chip QNN~\cite{HanruiWang2022OnchipQT} & Specific  \\ \cline{3-4} 
\multicolumn{1}{|c|}{} &  & CompVQC~\cite{ZhiruiHu2022QuantumNN} & General  \\ \cline{2-4} 
\multicolumn{1}{|c|}{} & \multirow{2}{*}{Optimization mapping method} & ML-QCP~\cite{HongxiangFan2022OptimizingQC} & General  \\ \cline{3-4} 
\multicolumn{1}{|c|}{} &  & HA~\cite{SiyuanNiu2020AHH} & General  \\ \hline
\multicolumn{2}{|c|}{Pulse Level Control} & PAN~\cite{DingYongshan2022PANPA} & General \\ \hline
\end{tabular}
\end{table*}

\section{Different kinds of Variational Quantum Algorithm}
\label{sec3}

To make better use of the quantum acceleration characteristics of quantum machines and solve various problems faced by current VQA, we need to summarize and classify the existing VQA methods. As shown in the following table, gate-level VQA methods can be divided into two categories: improving accuracy and improving efficiency. Furthermore, on this basis, the two directions will adopt different methods to achieve the goal. After that, pulse level control will also be discussed.

\subsection{Improving Accuracy}
The first direction we will discuss is to improve the accuracy and at the same time ensure that the efficiency of VQA is not greatly affected or even improved. To improve the accuracy, there are generally two method/s: Noise Suppression and Supercircuit. These two methods try to improve the accuracy of VQA from different angles. At the same time, they do not conflict with each other and can be used in combination.\\

\subsubsection{Noise Suppression}
Currently, we are in the NISQ phase, where quantum operations suffer from a high error rate of $10^{-2}$ to $10^{-3}$, much higher than the CPU/GPU's $10^{-6}$, so effective Noise Suppression can significantly improve VQA accuracy.

\textbf{QuantumNAT}
In this method, a noise suppression framework dedicated to PQC problems is proposed, which optimizes the robustness of PQC in both the training and reasoning phases.

In this method, the authors propose to collect the measured results of Noise-free Simulation and Real Device on a batch of input samples respectively for each qubit and calculate its mean value and standard deviation. The effect of quantum noise on PQC measurements can be observed as a linear mapping of noise-free measurements with scaling and shift factors. Therefore, the difference in feature distribution between noiseless and noisy scenes can be reduced by normalization. On the other hand, according to the actual noise model of quantum hardware, the author inserts quantum error gates into the PQC, to inject noise into the training process. Finally, post-measurement quantization is introduced to quantize the measurement results into discrete values to achieve the denoising effect.

It can be seen from the results that QuantumNAT experiments with four different QNN architectures on eight tasks running on five quantum devices. QuantumNAT always achieves the highest accuracy in 26 benchmark tests. On average, normalization, noise injection, and quantization improve the accuracy by 10\%, 9\%, and 3\% respectively. However, this method is only effective for PQC tasks and is hardware-dependent.

\textbf{QF-RobustNN}
It is a method to learn qubit errors in quantum networks through training. It proposes a general training framework for error perception learning, to reduce the impact of noise on neural networks.

As a result, the performance of QF-RobustNN was evaluated on IBM Qiskit-based simulators and IBM quantum processors. The accuracy of QF-RobustNN is much better than that of the general model under the condition of high error rate. On the other hand, QF-RobustNN has fewer additional SWAP gates and is more efficient than other methods due to the selection of appropriate mapping methods.

QF-RobustNN implements the learning of noise so that errors can be predicted, effectively reducing the impact of noise on QNN and improving the accuracy of the model. At the same time, due to the use of a more suitable mapping mode of quantum circuits, the number of additional gates can be effectively reduced, thus reducing consumption and ensuring efficiency. This method is an effective method for QNN, which is verified by simulators and tested by IBM Quantum Processor (i.e., the ibmq-montreal backend with 27 qubits). Due to the small number of machines used, can't determine whether the method is hardware-dependent.

\textbf{Summarize of Noise Suppression: }From the above two noise-aware methods, we can find that this method has an obvious effect in improving the accuracy of the model, but it doesn't significantly reduce the efficiency of the model, which is a fruitful attempt. However, most noise-aware methods need to learn the noise to achieve the effect of suppression, and the noise of different quantum devices or the same device at different times is always different, so such methods are generally hardware-dependent.

\begin{figure}[htbp]
	\centering
	\includegraphics[width=\linewidth]{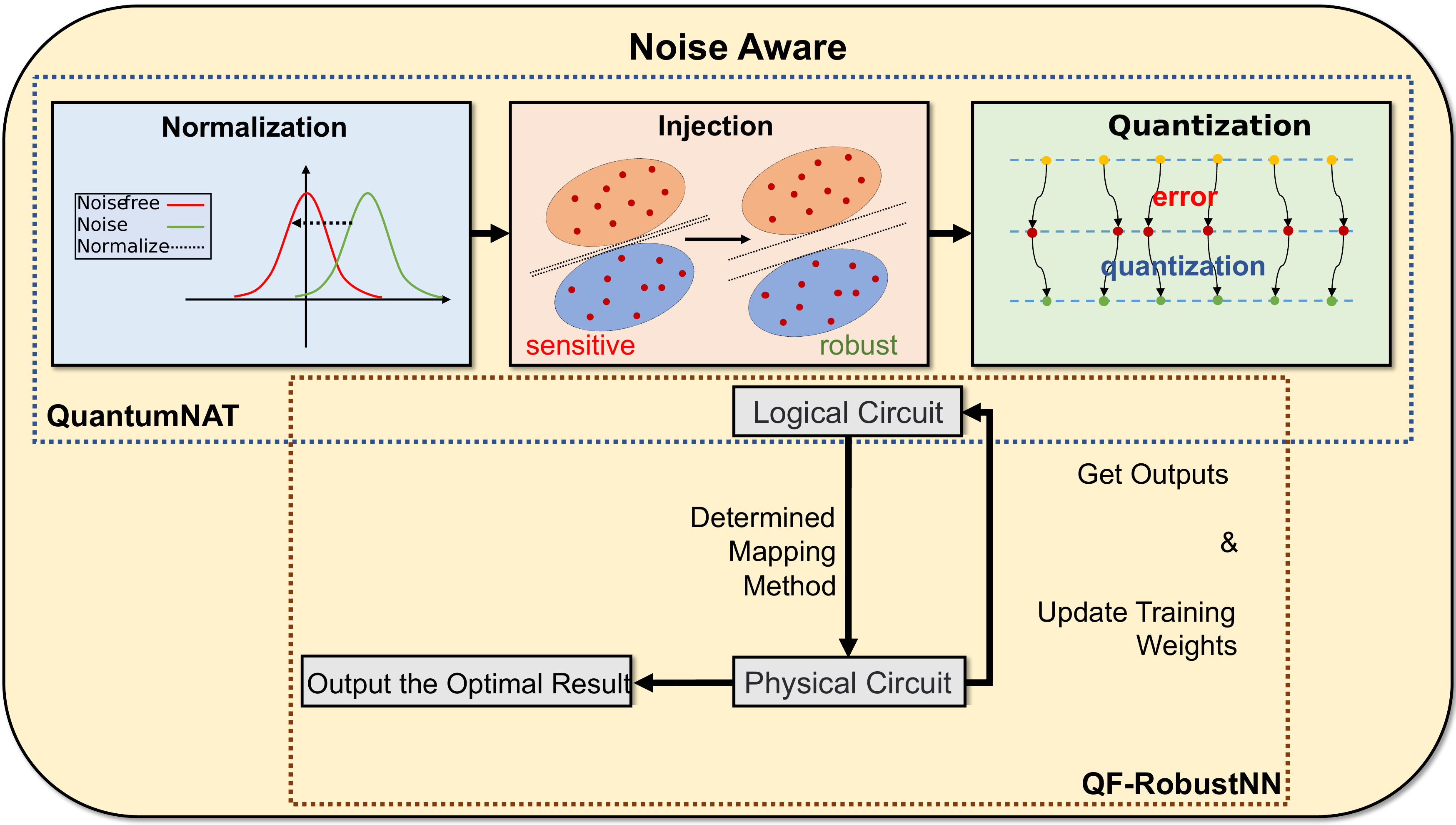}
	\caption{Workflow of commonly used noise-aware techniques implementation on quantum computing. The difference and correlation between works have been marked.}
	\label{fig:pic2}
\end{figure}

\subsubsection{Better Circuit}
Another way to improve accuracy is to find a circuit that is good enough to provide a relatively better circuit-qubit mapping for a given task on the target machine, an idea made possible by the adjustability of VQA parameters. In general, we can choose to build a super-circuit large enough to ensure that it contains the optimal solution we need, construct enough circuits and find the optimal one. Or, on the other hand, We can also choose to use some special architecture, or apply some methods from classical computers to this problem.

\textbf{QAS}
The current VQA is faced with two problems: 1. When the qubit number and circuit depth increase, the performance of vqa will decrease significantly due to the balance between expressiveness and trainability; 2. A deep circuit depth means that the gradient information received by the classical optimizer is full of noise, and the effective information disappears exponentially, which may lead to divergent optimization or barren plateaus. QAS solves the problem by unifying noise suppression and trainability enhancement into a learning problem.

QAS outperformed traditional VQAs in both machine learning and VQE tasks. In the task of data classification for quantum devices with noise, the classification accuracy of traditional quantum classifiers is only 50\%, while the quantum neural network and corresponding parameters output by QAS can reach 100\% accuracy.

\textbf{QuantumNAS}
 proposes noise-adaptive quantum circuits to provide a relatively better circuit-qubit mapping for a given task on a target machine. For parameterized quantum circuits, there can be many circuits with different architectures to achieve the same goal, and quantum gates with different numbers and positions are used. QuantumNAS can find the most suitable quantum circuit architecture for the target machine and the corresponding Qubit mapping.

This research draws on the idea of neural network search (NAS) in classical deep learning. First, construct a Supercircuit that is large enough. The Supercircuit contains multiple blocks, each with multiple layers of parameterized gates, whose parameters are accomplished by iteratively sampling and updating subcircuits. subcircuit is its subset. Due to the high training cost, only the gradient calculation of the subcircuit is carried out in the training of the supercircuit and the parameter subset of the supercircuit is updated. After building the right supercircuit, QuantumNAS uses genetic algorithms to find the right Subcircuit. Finally, redundant quantum gates are further removed to reduce the noise of the selected Subcircuit, and the iterative pruning method is used for pruning.

However, this method is hardware-dependent, because according to its experimental results, the accuracy of the experiment after three days' intervals is significantly reduced, which may be caused by IBM's correction machine.

\textbf{QMLP}
In this paper, we propose a quantum multilayer perceptron (QMLP) architecture, which enhances the fault tolerance of QNN circuits, provides abundant nonlinearity and proposes the use of a circuit with parameterized two-qubit entangled gates. The characteristics of this method are shown in that the fault-tolerant coding method provided by this method can generate more accurate low-level features, improve the original RUU layer design, provide adjustable nonlinearity, facilitate the construction of deep QNN, and ansatz constructed using parameterized double qubit gates.

The accuracy of QMLP in the ten-classification task on the MNIST dataset was 75\%, higher than that of QuantumFlow and QuantumNAS (69\% and 67\%). In the aspect of QNN design overhead, the number of QMLP gates and parameters are reduced by 2 times and 3 times respectively. However, as the circuit becomes more complex, the performance improvement will be offset by the side effects of gate errors when noise is involved, which shows certain limitations of this method. The tradeoff between model depth and noise tolerance should be considered when designing a QMLP model in practice.

\textbf{DRL-QAS}
This paper proposes a quantum architecture search framework with deep reinforcement learning (DRL) capability. In this framework, DRL agents have access to the expected value of PAUL-X, Y, Z and a predefined set of quantum operations for learning target quantum states, optimized by the dominant actor-Critical (A2C) and Proximal Strategy Optimization (PPO) algorithms. The advantage of this approach is that no quantum physics knowledge is encoded in the agent, and it can be used with other DRL architectures or optimization methods.

The authors demonstrate the successful generation of a multi-qubit GHZ state quantum gate sequence, proving that the method is indeed feasible, but the method requires a lot of resources, for the current NISQ phase, and still cannot be applied to the real machine.

\textbf{Summarize of Better Circuit: }The above methods are classified as a better circuit, which can also be divided into two different directions. Taking QAS and QuantumNAS as examples, the method constructs a supercircuit and selects a good enough subcircuit from it through their different methods. This method does play a role in improving accuracy. At the same time, this method is usually application general. However, due to the great workload of constructing the supercircuit, this part of the work still needs to be handed over to the classical computer in the NISQ environment. However, QMLP and DRL-QAS represent an attempt in another direction, which applies some methods on classical computers to the generation of an auxiliary circuit, to make the model stronger. The problem with this method is also reflected in the excessive workload of circuit construction, which can only be applied to the environment of the simulator at present.

\subsubsection{Approachs for particular problems}
In addition to more general methods, we can also find some specific methods for some special problems. Such attempts tend to be poor in scalability and elongation but can be very effective in improving the accuracy of individual problems.

\textbf{QUILT}
This method is aimed at quantum classifiers with multiple classification tasks, which often have poor accuracy. By establishing the integration of quantum classifiers QUILT makes up the deficiency of this field to some extent. A key idea behind the Quilt is the integration of smaller classifier models as opposed to using a single large classifier model. Such decisions ensure that the effects of quantum errors are minimized. This method divides classifiers into core classifiers that can classify all categories and OneVsALL classifiers that can improve the accuracy of each class. The problem is solved through the integration of one core classifier and multiple OneVsAll classifiers.

The performance of QUILT is equal to that of existing classifiers in binary tasks, but with the increase of categories, QUILT provides a 22\% and 46\% improvement in quad and octant tasks, respectively.

\subsection{Improving Efficiency}
The other direction we will discuss is to improve the efficiency of VQA while ensuring that its accuracy is not compromised. The reason for studying this direction is that quantum machines currently have fewer resources, and in the case of NISQ, it would be valuable to perform the same or more tasks with fewer resources.

\subsubsection{Pruning}
In the context of NISQ, the limited quantum resources limit the depth of the VQA. With appropriate pruning methods, removing circuits with little impact on the model can effectively compress the required resources at the expense of very little accuracy, thus making it possible to further increase the depth of the model.

\textbf{On-chip QNN}
In this context, the authors observed that in the vast majority of cases, small gradients tend to have large relative variations or even wrong directions under quantum noises, So such small gradient calculation is often unnecessary. By calculating only the gradients with high reliability and cutting out the small gradients, the resources required by the model can be compressed and the accuracy can be maintained or even increased.

The computational training engine of the In-situ gradient consists of three parts: 1. Jacobian calculation via parameter shift 2. Down-stream gradient backpropagation. 3. Gradient calculation. Finally, we can get the gradient of each parameter.

The author has conducted extensive experiments on 5 classification tasks with 5 real quantum machines, and the results are slightly stronger than that of fixed pruning. It can be seen that pruning does not affect the accuracy of the model, and the accuracy of the classification tasks of 2 and 4 kinds of images is more than 90\% and 60\%, respectively. Probabilistic gradient pruning is 7\% more accurate than unpruned QNN. Is the percent of the time and save $r*\frac{wp}{wa+wp}*100\%$, increased the efficiency. This method can only be applied to QNN, which has some limitations and can be implemented on real machines.

\textbf{CompVQC}
Due to the high noise and limited resources of NISQ devices, CompVQC proposed a quantum neural network pruning compression method based on ADMM.

The method consists of three phases. the first stage is preparation, starting with QNN pruning, which provides the basis of what pruning is in a quantum scenario; The authors propose that: A quantum gate can be pruned not only when its parameter/weight is 0. Then there is QNN Quantization,  some special values should be substituted to make its parameter/weight become 0, to reduce the depth. Finally, we need to build a Compression-Level look-up table (LUT) for each door. The second stage is Compression, which turns the non-abrupt problem with constraints into an unconstrained convex problem. Using ADMM to construct Compression Level LUT Reconstruction, to achieve QNN Compression. Finally, the author develops a logical quantum circuit that converts the above-compressed VQC input to the compiler for mapping into a fixed quantum device.

From the experimental results, CompVQC reduced circuit depth (almost more than 2.5\%), achieved a speedup of about 2.5 times, and reduced accuracy (<1\%) is negligible. This method is limited to VQC tasks and has poor scalability. However, the results of this method are similar on different quantum machines, which proves that it is not hardware-dependent.

\textbf{Summarize of Pruning: }Pruning is a very effective method to save resources and improve efficiency in the NISQ environment. The model efficiency of both On-chip QNN and CompVQC has been greatly improved. In addition, the accuracy of the model can be further improved by the correct pruning method to remove some counterproductive parts in the model. However, in terms of universality, this approach is often Device dependent, because different parts of the model may have different functions under different quantum devices.

\begin{figure}[t]
	\centering
	\includegraphics[width=\linewidth]{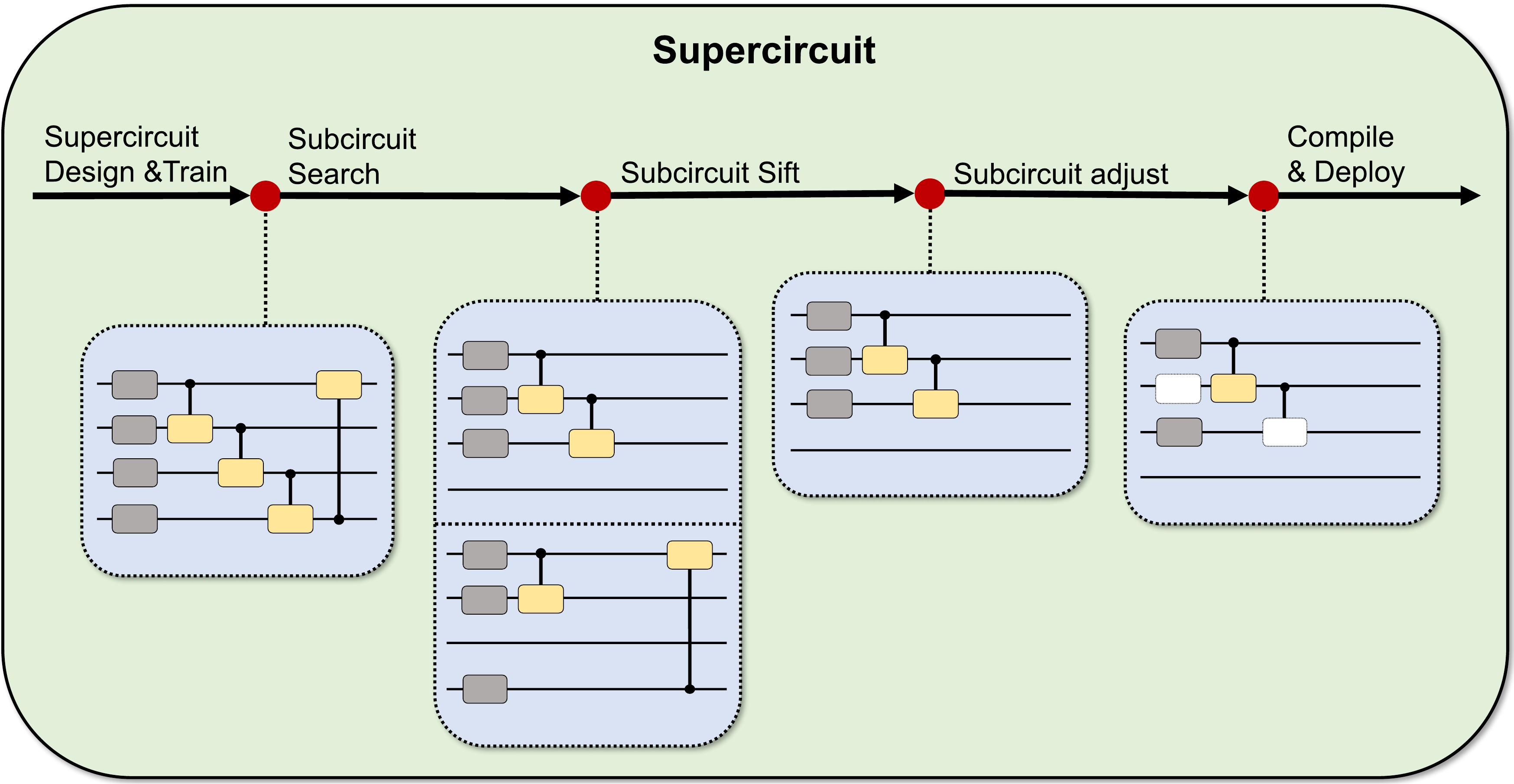}
	\caption{Overview of supercircuit. 1) Build and train the supercircuit according to the different algorithms. 2) Search for suitable subcircuits. 3) Sort the obtained subcircuits and select the best circuit. 4) Iterative trimming and fine-tuning to remove the excess gate. 5) Compile and deploy on the actual device.}
	\label{fig:pic3}
\end{figure}

\subsubsection{Optimization mapping method}
Due to the need to map logical qubits to physical qubits, this operation is not a one-to-one correspondence and often needs to convert complex gates in logical qubits into multiple basic gates~\cite{hua2022cqc, tan2020optimal, li2020towards, zulehner2018efficient}. Therefore, qubit mapping requires certain strategies to reduce the error rate and speed up the running of quantum programs.

\textbf{ML-QCP}
This paper describes QCP as a two-layer optimization problem and proposes a new framework based on machine learning (ML). This method uses a strategy-based deep reinforcement learning (DRL) algorithm for low-level combinatorial optimization problems. This method overcomes the disadvantages of the traditional heuristic method and the precise method, and at the same time can ensure the optimality-runtime trade-off. In this method, the low level is a combination problem, which is solved by deep reinforcement learning (DRL), while the high-level discrete search problem is solved by an evolutionary algorithm.
According to different tasks, the number of SWAP gates required by this method is reduced compared with other existing methods, and the maximum is more than 90\%. Optimize the circuits mapped to the machine by reducing the number of base gates to reduce consumption and improve performance, reducing the uptime by 40\%. However, this method may not be suitable for large-scale circuits and has some limitations.

\begin{figure}[t]
	\centering
	\includegraphics[width=\linewidth]{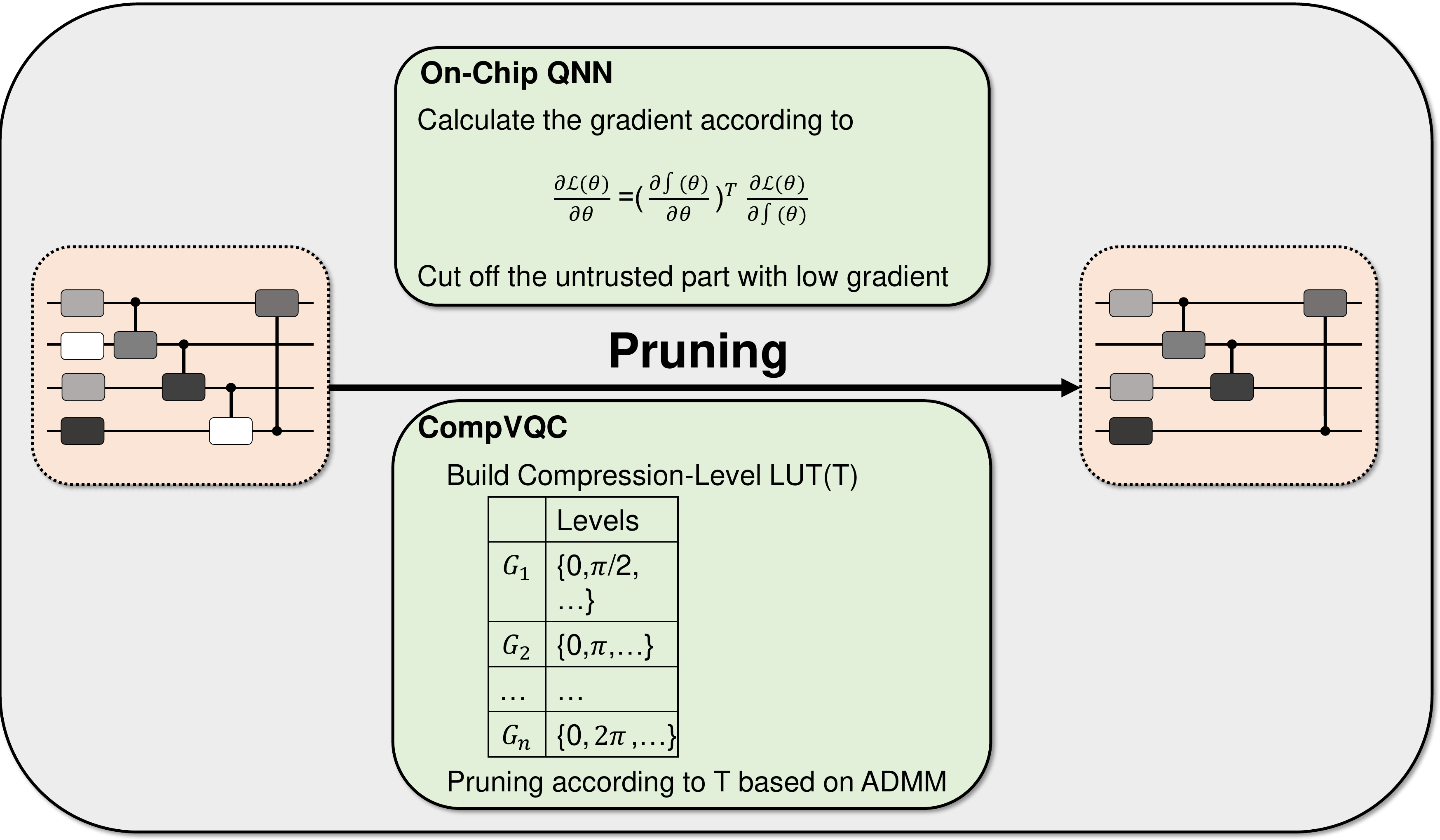}
	\caption{Two different ways for pruning: 1) Pruning according to the different gradients. 2) Establishing a Compression-Level LUT as the basis and pruning based on the ADMM.}
	\label{fig:pic4}
\end{figure}

\textbf{HA}
In this paper, a mapping conversion algorithm considering hardware topology and calibration data is proposed to improve the overall output state fidelity and reduce the total execution time. This method is based on the SABRE algorithm. By defining the appropriate heuristic function, the iterative method is used to ensure that all circuits get the appropriate mapping. In particular, in addition to choosing to insert SWAP gates, this method inserts Bridge gates if the topology allows.

The author used ibmq\_almaden and ibmq\_tokyo to compare the algorithm with the most advanced algorithms in terms of the number of additional gates. The optimal output was obtained when the HA algorithm was combined with the SABRE algorithm. Compared with the original circuit, the number of gates was reduced by 30\% on average, and the effect was better than using the SABRE algorithm alone. The authors claim that this method can be used on machines outside IBM, so it is not hardware dependent. The HA algorithm only takes into account calibration data, including gate errors and execution time. However, other physical constraints, such as crosstalk errors, may be included to account for crosstalk coupling between interconnects.

\textbf{Summarize of Optimization Mapping Method: }From the two examples of ML-QCP and HA, there is no doubt that a more effective mapping method will greatly reduce the consumption of resources and improve the efficiency of the model. At the same time, fewer basic gates also mean less noise, which can also provide an improvement in accuracy. Because of the transformation from logical to physical circuits during model deployment, the improved mapping method is probably the closest approach to the basics at the gate level, This approach is application-general and often hardware independent. (As shown in Fig. \ref{fig:pic4})

\subsection{Pulse Level Control}
\begin{figure}[htbp]
	\centering
	\includegraphics[width=\linewidth]{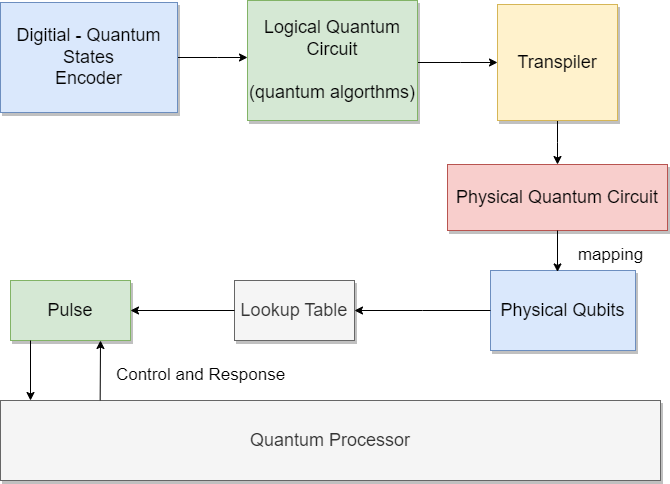}
	\caption{Compilation workflow of quantum computing.}
	\label{fig:work}
\end{figure}

The existing quantum algorithms basically take gate as the basic element, but the circuit construction in this way is not compatible with the underlying topology of quantum devices. Because in this case we need a compiler to compile it into a control signal on a physical qubit as shown in Fig. \ref{fig:work}. So there is a lot more room for improvement in the decomposition and tuning of the circuit at the pulse level~\cite{liang2022variational, liang2022pan, meitei2021gate, earnest2021pulse, magann2021pulses, ibrahim2022pulse,gokhale2020optimized, cheng2020accqoc}.

\textbf{PAN}
It was the first to demonstrate the feasibility of native-pulse ansatz on a NISQ machine. This avoids the redundancy that gate compilers can create, bypasses the severe limitations that exist with existing pulse generator optimizations such as QOC, and provides greater freedom.
PAN will first extract the configuration of the NISQ computer, including information about the quantum bit frequency and the mapping between the native gate and the native pulse. Next, build the native pulse ansatz from this information. Ansatz has two different types of layers, one consisting of single qubit primary pulses on all qubits, while the other consists of double qubit primary pulses on all connections, and both types of layers are inserted in turn during training. When building ansatz, you need to consider setting up entangled qubits with the topology of the corresponding NISQ machine to save on the consumption of qubit mapping. In particular, PAN proposes an incremental approach to extend the pulse circuit to limit the number of parameters held by the optimizer to ensure the optimizer's efficiency. After Ansatz is built, further pruning is needed to remove redundant parameters.
The results show that the effect of PAN is comparable to that of full configuration interaction, and the comparison with true amplitude Ansatz proves that PAN can reduce the duration by 73.6\% while maintaining the performance of Ansatz. From the above performance, pulse level operation is indeed superior to the general gate level operation in many aspects, because the pause level is more suitable for the characteristics of quantum devices and can provide greater training freedom. Compared with the gate level, there are fewer studies in this direction, so there are fewer references available in the construction of the model, which may be the difficulty of the attempt.

\section{Related Work}
In the quantum computing field, there has been a lot of existing work summarizing methods for better designing ansatz of VQAs.

~\cite{lu2022survey} talked about the error problem of QML and emphatically discussed how to deal with the Noise-aware problem of QML. However, the QML improvement Method discussed in~\cite{lu2022survey} is relatively simple compared with ours. Besides Noise Suppression, our paper also adds a summary and discussion of VQA improvement methods in other directions such as Pruning, Mapping methods, and Better Circuits. Non-general methods for special problems are also studied. Beyond the gate level, the pulse level is also an important research direction.

The main research object of~\cite{garcia2022systematic} and~\cite{mishra2021quantum} is QML, and they respectively summarized different algorithms in QML and their applications. Their content is limited to QML but with more details. While our paper discusses VQAs that include QML with details techniques of how to better design a good ansatz for VQAs.

\section{Conclusion}
Variational Quantum Algorithms are an important method of applying Quantum Computing in the NISQ environment. However, VQA is still facing many difficulties and further improvement is needed. This paper summarizes the existing VQA improvement approaches and outlines the different directions for improvement at the gate level and pulse level respectively. At the gate level, this paper first divides the methods into two categories for accuracy improvement and efficiency improvement further subdivides them and summarizes and evaluates different subdivision methods. This paper also discusses beyond circuit level work, which is on the pulse level. pulse level is a relatively new direction, but because of its characteristics, there is a lot more place for improvement in the decomposition and tuning of the circuit at the pulse level. Among the methods described here, some are computationally expensive, while others are exponentially dependent on quanta, and although they perform well in their environments, they are not suitable for large-scale quantum applications. Even though there are a lot of technologies out there, I think it's still an open question how to develop a method that is computationally cheap and can be implemented with large-scale quantum algorithms on real machines.

\bibliographystyle{elsarticle-num}
\bibliography{ref}

\end{document}